\documentclass[prl,twocolumn,showpacs,preprintnumbers,amsmath,amssymb]{revtex4}

\usepackage{graphicx}
\usepackage{dcolumn}
\usepackage{bm}
\usepackage{natbib}
\usepackage{amsmath}
\usepackage{here}
\usepackage{color}

\usepackage[normalem]{ulem}	

\begin{document}
\preprint{}

\title{Imaging fractional incompressible stripes in integer quantum Hall systems}

\author{Nicola Paradiso}
\author{Stefan Heun}\email{stefan.heun@nano.cnr.it}
\author{Stefano Roddaro}
\author{Lucia Sorba}
\author{Fabio Beltram}
\affiliation{NEST, Istituto Nanoscienze--CNR and Scuola Normale Superiore, Pisa, Italy}
\author{Giorgio Biasiol}
\affiliation{Istituto Officina dei Materiali CNR, Laboratorio TASC, Basovizza (TS), Italy}
\author{L. N. Pfeiffer}
\author{K. W. West}
\affiliation{Department of Electrical Engineering, Princeton University, Princeton, New Jersey 08544, USA}
\date{\today}

\begin{abstract}
Transport experiments provide conflicting evidence on the possible existence of fractional order within integer quantum Hall systems. In fact integer edge states sometimes behave as monolithic objects with no inner structure, while other experiments clearly highlight the role of fractional substructures. Recently developed low--temperature scanning probe techniques offer today an opportunity for a deeper--than--ever investigation of spatial features of such edge systems. Here we use scanning gate microscopy and demonstrate that fractional features were unambiguously observed in \textit{every} integer quantum Hall constriction studied. We present also an experimental estimate of the width of the fractional incompressible stripes corresponding to filling factors 1/3, 2/5, 3/5, and 2/3. Our results compare well with predictions of the edge--reconstruction theory.
\end{abstract}

\pacs{73.43.-f, 72.10.Fk}

\maketitle

Can a two--dimensional electron gas (2DEG) with integer filling factor show fractional quantum Hall (QH) effect? This question was posed more than 20 years ago by Beenakker~\cite{Beenakker1990} and was motivated by transport experiments~\cite{Kouwenhoven1990,Chang1989} suggesting that an integer QH edge can behave as if composed by a set of independent fractional channels that can be selectively populated and detected. Such a behavior is at odds with the edge model first proposed by Halperin~\cite{Halperin1982} in which each integer Landau level in the bulk gives rise to a chiral--edge mode that has no internal structure and can be described in terms of single--particle physics. More recent models including electron--electron interactions can explicitly describe the emergence of fractional QH substructures, as shown for instance by Chklovskii \textit{et al.}~\cite{Chklovskii1992}. Despite a number of experimental and theoretical studies, the issue of fractional order within integer QH systems is still an open question. A number of experiments showed clear indications of fractional phases in constrictions, either in terms of fractional quantization of conductance~\cite{Kouwenhoven1990,Chang1989} or Luttinger--like non--linear features~\cite{Roddaro2003,Roddaro2004,Roddaro2005,Roddaro2009}, although even the simple problem of how an ideal integer edge can ``branch'' and give rise to fractional edges remains unclear. On the other hand, recent interferometry experiments~\cite{Ji2003} and out--of--equilibrium energy spectroscopy data~\cite{Altimiras2009} indicate that an integer edge can behave as a monolithic object and shows no clear evidence of an inner structure. Whether such dual behavior depends on the specific device structure or is intrinsic, remains an unanswered question with potential important implications on the behavior of integer QH constrictions, which constitute the basic building block of QH interferometers. Obtaining direct, unambiguous experimental indications is hindered by the fact that fractional features are often difficult to identify, owing to the unavoidable random variability in real devices: even simple fractional conductance quantization steps can be easily masked by disorder or resonances. 

In this Letter we investigate transport in a set of QH constrictions by using scanning gate microscopy (SGM)~\cite{Paradiso2010,Paradiso2011} and show that {\em all} devices we studied display clear fractional--QH features. This observation is made possible by the fact that our technique allows us to finely tune constriction geometry and average out device--specific fluctuations. Our SGM maps directly probe the width of the most relevant fractional incompressible stripes (IS), corresponding to filling factors 1/3 and 2/5, together with their particle--hole conjugates~\cite{Girvin1984,MacDonald1994} 2/3 and 3/5. In these experiments we brought two counter--propagating integer--edge channels into proximity by means of a quantum point contact (QPC) and used the biased SGM tip to tune backscattering. From the measurement of the transmitted current as a function of tip position we can extract spatially--resolved information on the edge structure. Furthermore, we show that these results make it possible to quantitatively test predictions of the edge--reconstruction theory.

\begin{figure}[tbp]
\includegraphics[width=\columnwidth]{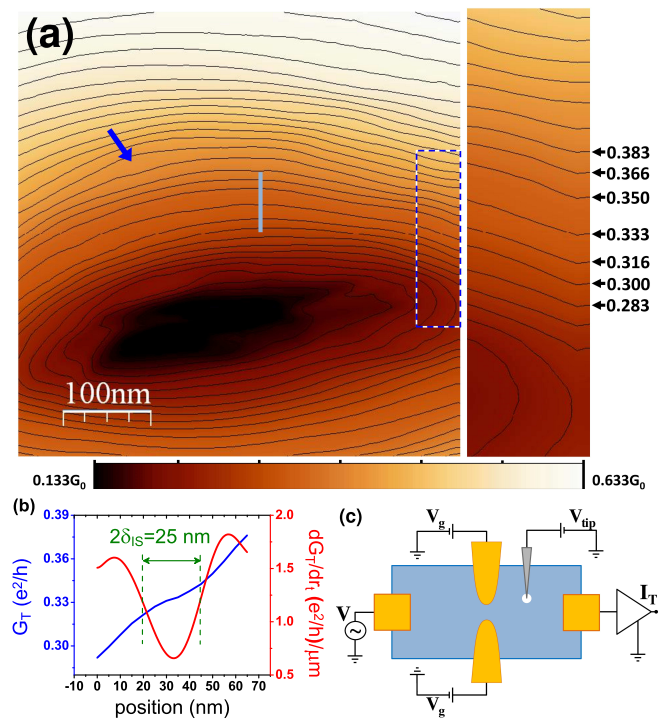}
\caption{(a) SGM  scan at the center of a QPC in a $\nu_b=1$ QH system ($V_{tip}=-6$~V). The map shows the transmitted differential conductance $G_T$ as a function of the tip position, together with contour lines at constant $G_T$.  The blue arrow indicates a structure induced by a local potential fluctuation. On the right, a zoom of the $50 \times 200$~nm region corresponding to the dashed rectangle is displayed. (b) Profile of $G_T$ along the light blue line in (a), together with its derivative. (c) Scheme of the SGM experimental setup.}
\label{fig:G094_FQHE}
\end{figure}

The configuration of our samples and the experimental setup is shown in Fig.~\ref{fig:G094_FQHE}(c). Hall bars were defined by standard optical lithography on AlGaAs--GaAs heterojunctions with an embedded two--dimensional electron gas (2DEG). The devices shown in this work were fabricated starting from three heterostructures: sample A has a 2DEG carrier density $n_A=1.77\times 10^{11}$~cm$^{-2}$ and a dark mobility $\mu_A=4.6\times 10^6$~cm$^{2}$/Vs. The corresponding values for samples B and C are $n_B=1.99 \times 10^{11}$~cm$^{-2}$, $\mu_B=4.5\times 10^6$~cm$^{2}$/Vs and $n_C=2.11 \times 10^{11}$~cm$^{-2}$, $\mu_C=3.88\times 10^6$~cm$^{2}$/Vs, respectively. The 2DEG depth is 80~nm for samples A and B, and 100~nm for sample C. QPCs were fabricated by thermal evaporation of Schottky split--gates (10~nm Ti/20~nm Au bilayer), defined by electron beam lithography.  The split--gate constriction gap was 300~nm wide for sample A and C, and 400~nm wide for sample B. Ohmic contacts were fabricated by thermal evaporation and annealing of Ni/AuGe/Ni/Au multilayers.

Scanning gate measurements were performed using the conductive tip of an atomic force microscope (AFM) to locally deplete the 2DEG via capacitive coupling.  The SGM is operated in a $^3$He cryostat with base temperature of 300~mK, while the electron temperature of the sample is about 400~mK, as measured by a Coulomb--blockade thermometer. 
Topographic scans were used to localize the QPC constrictions. To avoid shorts with the split--gates, the tip and the gates were kept grounded during this operation. After acquiring the topography of the relevant area, the tip was negatively biased and scanned about 40~nm above the sample surface.  SGM maps are obtained by measuring the transport signals at each tip position, as shown in Fig.~\ref{fig:G094_FQHE}(c). Data were processed with the WSxM software~\cite{Horcas2007}. More details can be found in~\cite{SI}.

Figure~\ref{fig:G094_FQHE}(a)  shows a SGM measurement on a QPC on sample B in the QH regime at bulk filling factor $\nu_b=1$ ($B=8.23$~T). The scan setup is similar to the one reported in Ref.~\cite{Paradiso2010}. The split--gate axis is vertical with respect to the scan area, so that the source--drain current flows horizontally. The scan is centered approximately 100~nm above the QPC center. The split--gate voltage is set to $V_g=-0.30$~V, which allows to set the filling factor under the gates to $\nu=0$ without inducing backscattering between the counter--propagating edges inside the constriction (transmission of the QPC $t=1$). Figure~\ref{fig:G094_FQHE}(a) is a map of the source--drain (transmitted current) differential  conductance $G_T$ as a function of the tip position, with a bias $V_{tip}=-6$~V applied to the tip. As clearly shown in the figure, when the distance $r_t$ of the tip from the QPC center is gradually reduced, backscattering is enhanced and $G_T$ is suppressed. This global trend is consistent with earlier measurements in the integer QH regime with $\nu_b\geq 2$~\cite{Aoki2005,Paradiso2010}. In these latter measurements, owing to the details of edge reconstruction, plateaus were observed whenever the tip induced an incompressible phase at the QPC center. The $G_T$ value inside the plateau regions was reported to be quantized to a multiple of the quantum of conductance $G_0\equiv e^2/h$. As pointed out in Refs.~\cite{Aoki2005,Paradiso2010}, the plateau width is approximately twice the width $\delta_{IS}$ of the IS in the reconstruction model. In fact, reducing $r_t$ by approximately $2\delta_{IS}$, the adjacent compressible stripes (CS) merge at the QPC center, so that further backscattering is induced. SGM was thus shown to be an effective tool to \textit{image} the IS.

If electron--electron interactions are taken into account, the reconstruction picture describes the edge in quantum Hall liquids at $\nu_b=1$ as a series of alternating CS and IS. These structures are difficult to detect with other scanning probe techniques but can be successfully revealed by SGM measurements, as we demonstrate in this Letter. On the right--hand side of Fig.~\ref{fig:G094_FQHE}(a), we show a $50$~nm$\times 200$~nm zoom of the region corresponding to the dashed rectangle. The contour line density plot presents a shoulder corresponding to a plateau for $G_T=G_0/3$. This plateau can be directly observed in Fig.~\ref{fig:G094_FQHE}(b), where we show the conductance profile acquired along the light blue line in Fig.~\ref{fig:G094_FQHE}(a), together with its derivative. From the half--width of the minimum in the derivative we can estimate the width of the fractional IS $\delta_{IS} \simeq 12$~nm.

\begin{figure}[tbp]
\includegraphics[width=\columnwidth]{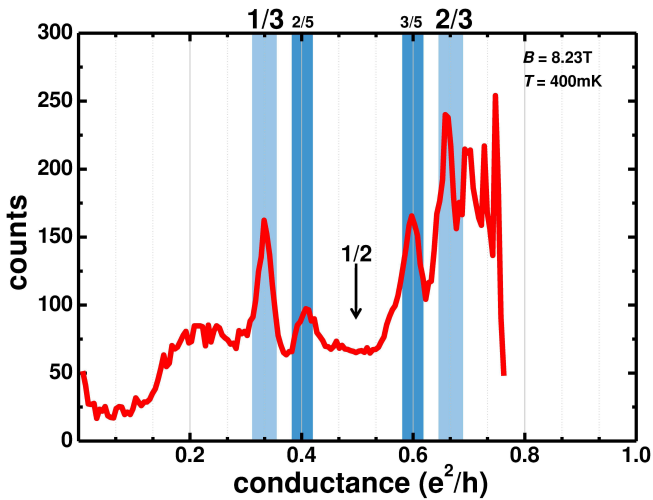}
\caption{Histogram of the average occurrence of each $G_T$ value within 9 different SGM scans performed at different $V_{tip}$ values. 
Peaks for $G_T=1/3$, 2/5, 3/5, and 2/3 are visible.}
\label{fig:FQHE_peaks}
\end{figure}

A straightforward way to highlight the presence of plateaus in the conductance is to count the number of times each value of $G_T$ occurs in the SGM map and plot these counts in a histogram~\cite{SI}. The presence of a plateau implies that the corresponding $G_T$ value is found more often in the SGM image; this in turn implies that a peak is produced in the histogram. 
Figure~\ref{fig:FQHE_peaks} shows the graph resulting from the averaging of 9 SGM scans performed on the same QPC at different tip voltages ($V_{tip}$ from $-7.5$ to $-3.5$~V). The histograms relative to the 9 individual scans are reported in the Supplemental Material~\cite{SI}. Peaks for $G_T= 1/3$, 2/5, 3/5 and 2/3 are clearly visible. Such values correspond to the most relevant fractions (1/3 and 2/5) together with their symmetry conjugates (2/3 and 3/5). Note that the observed number of fractional stripes is in agreement with the estimate provided by Chklovskii in Ref.~\cite{Chklovskii1995}.
The spurious peaks, visible for  $G_T$ values larger than 2/3, are caused by cut-off effects, which are discussed in the Supplemental Material~\cite{SI}. Similar measurements were performed on six samples, and at least the 1/3 peak was always unambiguously identifiable for all  different tip voltages. The amplitudes of the different peaks reflect the relative robustness of the fractions, e.g.~in Fig.~\ref{fig:FQHE_peaks} the 1/3 peak is three times larger than the 2/5 peak.  The averaging procedure makes it possible to effectively sample the whole conductance range from 0 to 2$G_0$/3. In fact, in scans with high $V_{tip}$ the higher $G_T$ values lie outside the scan area. Vice--versa, in scans with low $V_{tip}$, $G_T$ is higher than 2$G_0$/5 even at the QPC center. Hence, the averaging operation over many values of $V_{tip}$ is useful since it increases statistics, improves peak visibility with respect to fluctuations, which are averaged out further, and increases the range of $G_T$ values sampled in our analysis (i.e.~the range of the abscissa in Fig.~\ref{fig:FQHE_peaks}).

SGM scans can better highlight even weak structures since they provide much more information than a single sweep of the split--gate potential. Even though weak structures may not be readily recognizable in a single sweep, they become evident in a histogram graph, where all spurious structures are averaged out. 
This is demonstrated by the fluctuation indicated by the blue arrow in Fig.~\ref{fig:G094_FQHE}. As we show in the Supplemental Material~\cite{SI}, the SGM scan allows to average the $G_T$ signal on a large two-dimensional region, so that local fluctuations have a negligible effect on the global histogram.
We wish to stress that the ability to resolve thin stripes stems from the higher spatial resolution provided by the SGM technique compared to scanning force or scanning capacitance microscopy~\cite{Tessmer1998,McCormick1999,Yacoby1999,FinkelsteinScience2000,Weitz2000,Ahlswede2002}. In fact, here the resolution is only related to the accuracy of the piezo scanner that controls the tip--sample position, which is of the order of 0.1~nm. Even though the width of the electrostatic potential induced by the tip is relatively large (typically more than 100~nm), what matters here is how accurately the equipotential contour is moved, i.e.~the precision of the lateral displacement of the edge.

\begin{figure}[tbp]
\includegraphics[width=\columnwidth]{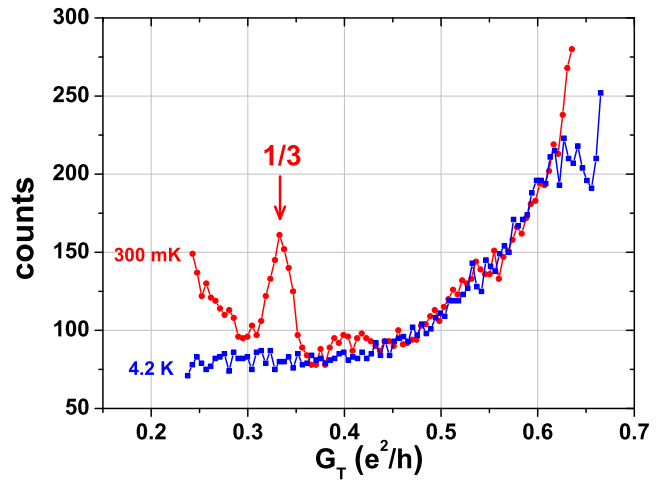}
\caption{Histogram plots of the occurrences of  $G_T$ values for two SGM scans performed at 300~mK (red curve) and 4.2~K (blue curve), on a QPC fabricated on sample C. The increase of temperature completely washes out the fractional IS, so that the 1/3 peak disappears.}
\label{fig:FQHE_temp}
\end{figure}

Figure~\ref{fig:FQHE_temp} shows the impact of  temperature on the visibility of fractional peaks, measured on a QPC fabricated on sample C. While at 300~mK a peak for $G_T=G_0/3$ is clearly observed, at a base temperature of 4.2~K the 1/3 peak completely disappears, and the curve becomes featureless. This is consistent with the picture of an incompressible stripe originating from the condensation of fractional quasi--particles with an excitation gap $\Delta_{1/3}$ of the order of 1~K  ($\approx$~100~$\mu$eV), as estimated from tunneling measurements on samples with similar characteristics~\cite{Roddaro2009}. This value is also consistent with recent magnetocapacitance experiments~\cite{Khrapai2008} that yielded a chemical potential jump across the fractional gap of the order of $\Delta\mu_{1/3}=3\Delta_{1/3}\approx 400$~$\mu$eV at 0.5~K. The fractional gap is rapidly suppressed as the temperature increases~\cite{Khrapai2008}, so that at 4.2~K virtually all the quasi--particles are excited, therefore screening is  effective and compressibility increases.

\begin{figure}[tbp]
\includegraphics[width=\columnwidth]{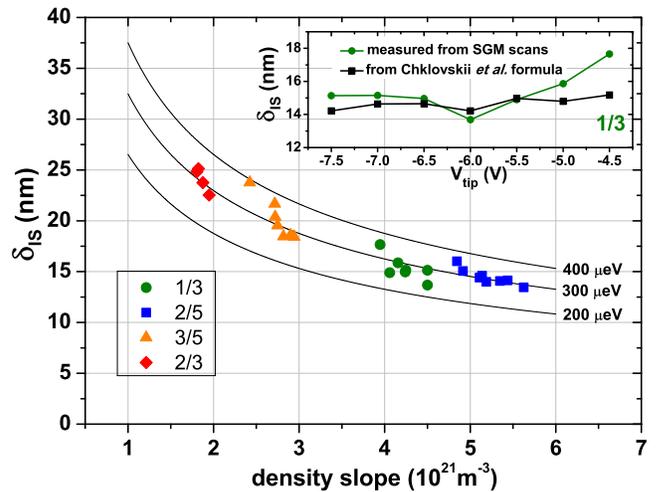}
\caption{IS width $\delta_{IS}$ plotted as a function of the electron density gradient (scatter plots), together with the reconstruction picture predictions for $\Delta\mu_f=200$, 300 and 400~$\mu$eV (thin lines). (inset) Plot of  $\delta_{IS}$ for the 1/3 fraction measured for different $V_{tip}$ values (green spots), together with the $\delta_{IS}$ values calculated with the formula of Chklovskii \textit{et al.}~(black squares). The latter have been determined by estimating the electron density gradient corresponding to each scan, and assuming that $\Delta_f=1$~K (see text).}
\label{fig:global}
\end{figure}

It is possible to estimate the range of $G_T$ values corresponding to a given plateau in the data of Fig.~\ref{fig:FQHE_peaks} by considering the full width at half maximum (FWHM) the corresponding peak. These intervals of $G_T$ correspond to a stripe in the SGM map, whose average width is a good approximation of the fractional plateau width. By applying this procedure to all SGM scans, we can extract the value of $\delta_{IS}$ of each fractional IS. These values are consistent with those obtained from a direct estimate of the plateau width, as the one shown in Fig.~\ref{fig:G094_FQHE}(b). In order to compare these values with the predictions of the reconstruction picture, however, it is necessary to estimate the local electron density gradient in correspondence of the IS, since in the reconstruction picture the square of the IS width is proportional to the energy gap between edge states and inversely proportional to the gradient of electron density~\cite{Chklovskii1992,Erratum}. SGM scans yield an estimate of the latter value by measuring the slope of $G_T$ near the plateaus. $G_T$ is proportional to the local filling factor induced by the tip potential at the QPC center, which in turn is proportional to electron density. Thus the slope in $G_T$ near the plateau is proportional to the electron--density gradient near the IS~\cite{SI}.
For each SGM scan we can thus compare this experimental density--gradient value to the $\delta_{IS}$ value expected from the formula of Chklovskii \textit{et al.}~\cite{Erratum} (inset of Fig.~\ref{fig:global}).  Both the absolute values and the trends of the reconstruction model predictions are in good agreement with the experimental data~\cite{SI}. This validates the analysis method and allows to convert tip--voltages into an universal electron density gradient scale. In the main panel of Fig.~\ref{fig:global} we report all measured $\delta_{IS}$ as a function of the electron density gradient. Data are shown together with the predictions of the formula of Chklovskii \textit{et al.}~\cite{Erratum} for $\Delta\mu_f=200$, 300, and 400~$\mu$eV. The agreement between the data and the reconstruction model is remarkable, especially in light of the uncertainty on the fractional--gap value, which is known to be rather sensitive to the details of disorder potential. Notably, data globally follow the expected $(dn/dr_t)^{-1/2}$ dependence.

These results convincingly demonstrate the occurrence of IS at sample edge when the filling factor equals the most robust fractions. Such IS are wider than the magnetic length ($\ell=(\hbar /eB)^{1/2}=9$~nm) and are able to isolate the different CS. This can explain why the fractional components behave as independent channels that can be selectively populated and detected~\cite{Beenakker1990,Kouwenhoven1990,Chang1989,Deviatov2006}.  The presence of fractional IS also explains the observation of Luttinger liquid behavior in tunneling experiments between $\nu=1$ phases (Fermi liquids), presented in Ref.~\cite{Roddaro2009}. Such results were interpreted by assuming that electrons tunnel through a region with local fractional filling factor $\nu^{\ast}$ separating the two main incompressible phases at $\nu=1$. The present work shows that such a region is precisely the fractional IS that is present at the sample edge. The QPC was thus used to individually partition the fractional components within an integer edge.

In conclusion, our spatially--resolved study of the edge structure sheds new light on quantum Hall physics and in particular on the complex phenomena recently reported in transport experiments~\cite{Bid2010}. In fact, the role of fractional phases in quantum interferometry is still not clear but this knowledge may open up exciting developments. For instance, the ability to controllably partition an integer edge and partially transmit one of its fractional components may be the key for the implementation of fractional quasi--particle Mach--Zehnder interferometers, currently one of the main goals in the field of coherent quantum transport.

\begin{acknowledgments}
We acknowledge financial support from the Italian Ministry of Research (MIUR--FIRB projects RBIN045MNB and RBID08B3FM).
\end{acknowledgments}

\end{document}